\documentclass[apj]{emulateapj}
\lefthead{TSUJIMOTO \& BEKKI }
\righthead{GC FORMATION  FROM SN IA EJECTA}

\def\ltsima{$\; \buildrel < \over \sim\;$}
\def\ltsim{\lower.5ex\hbox{\ltsima}}
\def\gtsima{$\; \buildrel > \over\sim \;$}
\def\gtsim{\lower.5ex\hbox{\gtsima}}
\def\ms{$M_{\odot}$ }
\def\msp{$M_{\odot}$}

\slugcomment{Accepted for publication in ApJ Letters}

\begin{document}
\title{First Evidence of Globular Cluster Formation from the Ejecta of Prompt Type Ia Supernovae}

\author{Takuji Tsujimoto\altaffilmark{1,2} and Kenji Bekki\altaffilmark{3}}

\affil{$^1$National Astronomical Observatory of Japan, Mitaka-shi,
Tokyo 181-8588, Japan; taku.tsujimoto@nao.ac.jp \\
$^2$The Graduate University for Advanced Studies, Hayama, Kanagawa 240-0193, Japan \\
$^3$ICRAR, M468, The University of Western Australia, 35 Stirling Highway, Crawley Western Australia 6009, Australia
}

\begin{abstract}

Recent spectroscopic observations of globular clusters (GCs) in the  Large Magellanic Cloud (LMC) have discovered that one of the intermediate-age  GC, NGC 1718 with [Fe/H]=-0.7 has  an extremely 
low [Mg/Fe] ratio of $\sim$-0.9. We propose that NGC 1718 was formed from the ejecta of type Ia supernovae (SNe Ia) mixed with  very metal-poor ([Fe/H] $<$-1.3) gas about $\sim 2$ Gyr ago. The proposed scenario is shown to be consistent with the observed abundances of Fe-group elements such as  Cr, Mn, and Ni. In addition, compelling evidence for asymptotic giant branch stars playing a role in chemical enrichment during this GC formation is found. We suggest that the origin of the metal-poor gas is closely associated with the efficient gas-transfer from the outer gas disk of the Small Magellanic Cloud to the LMC disk. We anticipate that the outer part of the LMC disk contains field stars exhibiting significantly low [Mg/Fe] ratios, formed through the same process as NGC 1718.

\end{abstract}


\section{Introduction}

Chemical abundance of stars is a fossil record on how the stars are formed. Thus we can assess the origin of a globular cluster (GC),  which still remains unresolved in spite of many previous observational and theoretical efforts, if we detect some signal characteristic of a specified nucleosynthesis result in the abundances of its member stars. Recently, the GC in the Large Magellanic Cloud (LMC), NGC 1718, is found to exhibit [Mg/Fe]=-0.9$\pm$0.3 from high-resolution spectra \citep{Colucci_12}. This is indeed a surprising result since no stars with such a low ratio have been found in the Galaxy or the LMC for field or star clusters. Only two stars in the Carina dwarf spheroidal (dSph) galaxy has been detected so far as the similar level of deficiency of Mg ([Mg/Fe]= -0.73$\pm$0.41, -0.95$\pm$0.33) for [Fe/H]=-1.2 in the field \citep{Lemasle_12}.

It is well-known that the Galactic GCs show  the Al-Mg and O-Na anticorrelations and accordingly contain stars with low Mg and O abundances in GCs  \citep[e.g.,][]{Gratton_04}. However, the Al-Mg abundance range is far smaller than the variation in the Na-O abundances, setting the lowest [Mg/Fe] ratio at $\sim$-0.3 \citep[e.g.,][]{Sneden_04}. In addition, a lack of such anticorrelations in young and intermediate-age GCs in the LMC has been reported \citep{Mucciarelli_11}. Figure 1 shows the observed [Mg/Fe] ratios in the LMC together with in the Galaxy for both GCs and field stars. For GCs which are observed for more than two stars, the [Mg/Fe] ranges are shown for individual GCs. It is evident that the location of the abundance of NGC 1718 in the [Mg/Fe]-[Fe/H] diagram is unusual.

NGC 1718 has an age of $\sim$2 Gyr \citep{Grocholski_06} and [Fe/H]=-0.7 \citep{Colucci_12}, the lowest among the intermediate-age LMC clusters (a mean [Fe/H] of $\sim$-0.5). In fact, its very low [Mg/Fe] ratio of $\sim$-0.9 together with these two properties give us the theoretical basis to decipher the origin of this cluster. First of all, such a very low [Mg/Fe] is considered to be exclusively associated with the nucleosynthesis product  from type Ia supernovae (SNe Ia). Nucleosynthesis in SNe Ia gives [Mg/Fe]$\approx$-1.5 \citep{Iwamoto_99}, while individual type II SNe (SNe II) including hypernovae seem hard to predict as small as [Mg/Fe]$<$-0.2 \citep[e.g.,][]{Kobayashi_06}. Though some low-mass SN II models predict very low [Mg/Fe] ratios down to $\sim$-0.6 \citep[see][]{Gibson_97}, no imprints of these SNe II are seen in the chemical abundances of Galactic halo stars. Therefore, it is expected that [Mg/Fe]=-0.9 can be explained by the hypothesis that the stars are born from the ejecta of SNe Ia mixed with a surrounding interstellar matter (ISM) with a moderate [Mg/Fe] ratio \citep{Tsujimoto_06}.

However, at the formation epoch of NGC 1718, i.e., $\sim$2 Gyr ago, an ISM in the LMC is likely to be enriched up to [Fe/H]$\sim$-0.5. When SNe, whether II or Ia, explode in such an enriched ISM,  relic of nucleosynthesis in individual SNe is hardly imprinted in stellar abundances since they basically reflect chemical abundances of the ISM \citep{Tsujimoto_98}. Accordingly, SN Ia explosion in a low metallicity ISM is crucial for the formation of stars with their elemental features characteristic of SN Ia nucleosynthesis. Then, a key question here is how the low metallicity ISM required for this GC formation was obtained in the LMC that had been already enriched to [Fe/H]$\sim$-0.5 about 2 Gyr ago.

The LMC has a unique history that is different from the Galaxy through the interactions with the Small Magellanic Cloud  (SMC) and the Galaxy. \citet{Bekki_07} and \citet{Diaz_12} reveal that the LMC has a close encounter with the SMC about 2 Gyr ago, which leads to disrupt the disk of the SMC, create the Magellanic Stream (MS), and eventually accrete the gas from the SMC onto the LMC by a total mass of $\sim 10^8$\msp. Since the outer gas-rich disk of the SMC is a major source of the accreting gas, its metallicity is expected to be very low. If we take a look at  the abundance of the MS, \citet{Fox_10} deduces [O/H]=-1.0$\pm$0.13. For the Magellanic Bridge (MB) which share the same origin with the MS, \citet{Rolleston_99} measure the abundances of young stars resided in the MB, and find $\sim$ 1 dex lower abundance than the solar value on average, including [Si/H]=(-1.23 - -1.46)$\pm$0.25 for one star. Moreover, \citet{Misawa_09} claims a large inhomogeneity of the observed chemical composition within the MB gas. Thus, the accretion of a metal-poor gas initially belonging to the SMC could involve gas with a sufficiently low metallicity to meet the condition  for this GC formation.

In this {\it Letter}, we present a new  scenario where the GC NGC 1718 was formed from a metal-poor gas mixed exclusively with SNe Ia ejecta about 2 Gyr ago. In this scenario,  the metal-poor gas initially in the outer part of
the SMC was accreted onto a part of the LMC disk, some regions of which were devoid of chemically enriched ISM like the HI holes currently observed in the LMC disk \citep{Kim_99}.

We start with the estimate of metallicity and mass of the ISM swept-up by one single SN Ia ejecta, and then the total number of SNe Ia, required for explaining  the observed properties of NGC 1718. Based on these results, a new scenario for the formation of NGC 1718 is presented in detail. Subsequently, we demonstrate  that our scenario is consistent with the observed abundances of other elements  such as Fe-group and light-odd elements.

\begin{figure}[t]
\vspace{0.2cm}
\begin{center}
\includegraphics[width=7cm,clip=true]{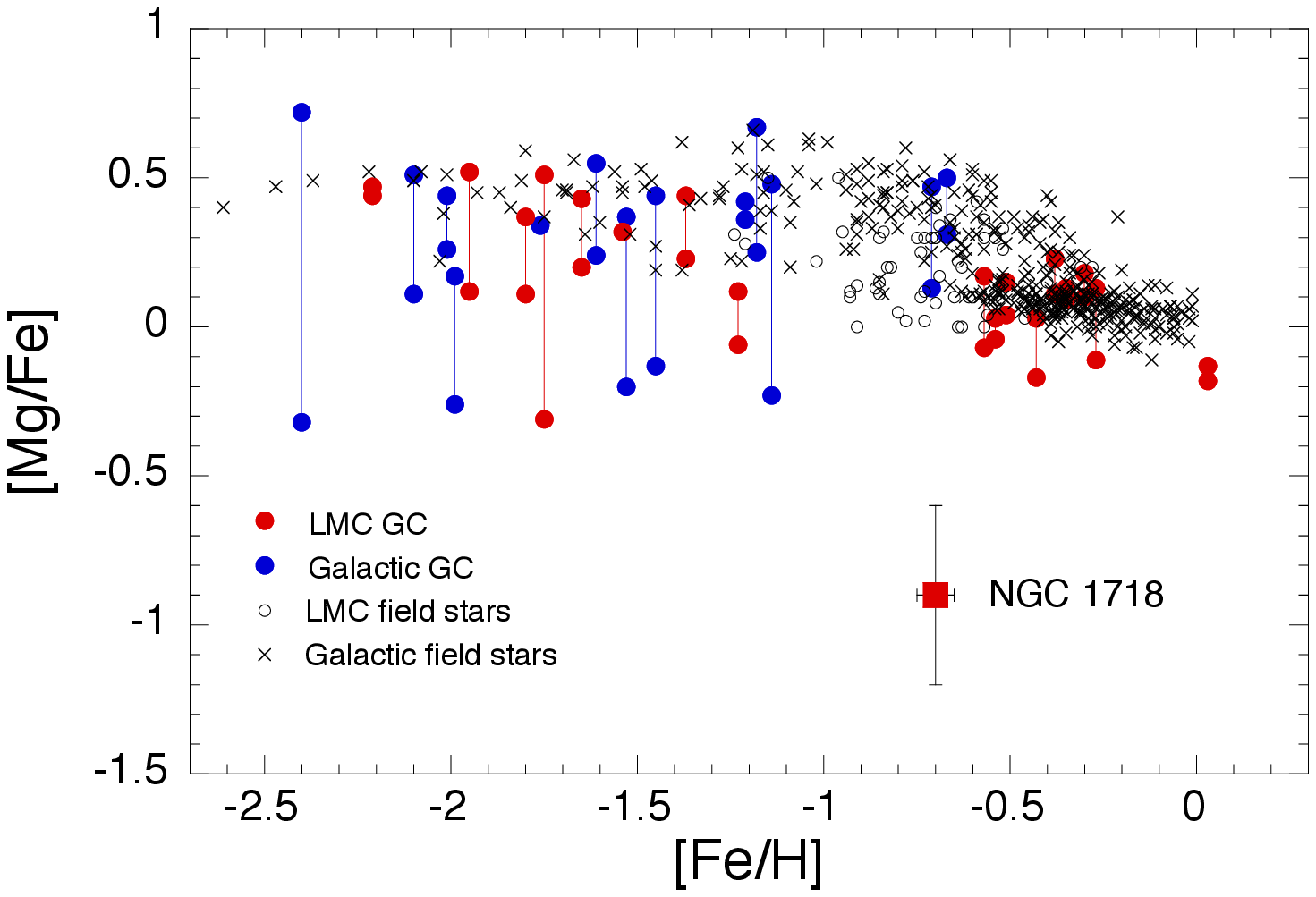}
\end{center}
\vspace{0.3cm}
\caption{Observed [Mg/Fe]-[Fe/H] correlation for both the LMC and the Galaxy. The data of LMC clusters (red filled circles) are taken from \citet{Johnson_06}, Mucciarelli et al.~(2008, 2010, 2011), and \citet{Colucci_12} together with NGC 1718 plot with an error bar \citep[filled red square:][]{Colucci_12}. The Galaxy GC data are shown by blue filled circles \citep[objects and references are from Table~8 in][]{Mucciarelli_10}. The variation in [Mg/Fe] for individual GCs are shown by bars. Field stars are denoted by open circles for the LMC \citep{Pompeia_08} and crosses for the Galaxy \citep{Gratton_03, Reddy_03, Bensby_05}.
}
\end{figure}

\section{How to make a very low [Mg/Fe] ratio}

\begin{figure}[t]
\vspace{0.2cm}
\begin{center}
\includegraphics[width=7cm,clip=true]{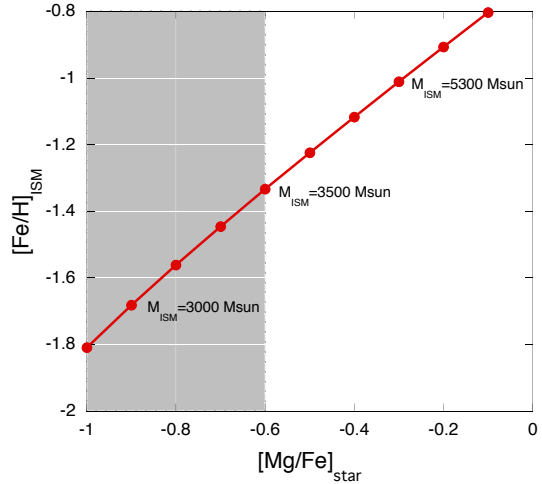}
\end{center}
\vspace{0.3cm}
\caption{The predicted relation between the metallicity of the ISM, [Fe/H]$_{\rm ISM}$, that is mixed with the ejecta of SN Ia and the abundance ratio inherited to a star, [Mg/Fe]$_{\rm star}$, that is born inside the ejecta. The calculated mass swept-up by a single SN Ia, $M_{\rm ISM}$, is denoted for three cases corresponding to [Mg/Fe]$_{\rm star}$=-0.9, -0.6, and -0.3.
}
\end{figure}

In this section, we estimate the [Mg/Fe] ratio of the ISM mixed with the ejecta from a single SN Ia as a function of the metallicity ([Fe/H]$_{\rm ISM}$) and the total mass of the ISM ($M_{\rm ISM}$). The [Mg/Fe] ratio thus obtained corresponds to the stellar [Mg/Fe] ([Mg/Fe]$_{\rm star}$) for the stars formed from the final ISM.
A SN Ia ejects the masses of 0.63 \msp, 8.5$\times 10^{-3}$\ms of Fe and Mg, respectively \citep{Iwamoto_99}. These heavy-elements are assumed to be well mixed with the original ISM with [Fe/H]$_{\rm ISM}$ and $M_{\rm ISM}$. From the mixes gas, new stars are eventually born with the chemical feature of [Mg/Fe]$_{\rm star}$ and [Fe/H]=-0.7, that is equivalent to the metallicity of NGC 1718. Here we adopt  [Mg/Fe]=0 for a reasonable value of the original ISM, because we  consider the situation where the ISM is enriched owing to a long-term chemical evolution until $\sim$2 Gyr ago, and thus must involve the sufficient contribution from SNe Ia.

Based on simple analytical calculations, we find the combination of  [Fe/H]$_{\rm ISM}$ and $M_{\rm ISM}$ required for the formation of stars with [Fe/H]=-0.7 and a given [Mg/Fe]$_{\rm star}$ ratio. 
The result is presented in Figure 2, showing the relation between [Mg/Fe]$_{\rm star}$ and [Fe/H]$_{\rm ISM}$. 
The values of $M_{\rm ISM}$ are  assigned  to three cases for [Mg/Fe]$_{\rm star}$ =-0.9, -0.6, and -0.3. 
The shaded region corresponds to the range of [Mg/Fe] ratio for NGC 1718 including an observed error \citep{Colucci_12}. The obtained relation crossing the shaded range leads to the conclusion that a very low [Mg/Fe] observed in NGC 1718, i.e., [Mg/Fe]$\leq$-0.6, can be achieved only if the metallicity of the ISM mixed with the SN Ia ejecta is  lower than [Fe/H]$_{\rm ISM}\sim$-1.3. On the other hand, an enriched ISM with the metallicity such as [Fe/H]$_{\rm ISM}\sim$ -0.5 as expected at $\sim2$  Gyr ago in the LMC will completely erase information on the nucleosynthesis of SN Ia retained inside its ejecta.

\section{Proposed scenario}

From the result on the required two properties of the ISM mixed with the ejecta of SN Ia, i.e., [Fe/H]$_{\rm ISM}<$-1.3 and $M_{\rm ISM}\sim$3000 - 3500\msp, the following two conditions should be set to build NGC 1718 as observed; (i) The ISM that had been already present in the LMC disk was mostly expelled from the vicinity of future NGC 1718 before the event of a low-metallicity gas accretion to make a new ISM with [Fe/H]$<$-1.3, (ii) tens of SN Ia exploded to have the observed  mass of the GC. 

First, we start to discuss the issue on a mass budget. The mass of NGC 1718 is estimated to be 6$\times 10^4$\ms \citep{Mackey_03}. Then, it turns out that sequential $\sim$17-20 SNIa explosions should occur.  Here we should highlight recent results regarding the delay time distribution (DTD) of SNe Ia yielded by the studies on the SN Ia rate in distant and nearby galaxies. These studies dramatically shorten the SN Ia's delay time, compared with its conventional timescale of $\sim$1 Gyr \citep{Pagel_95, Yoshii_96}. \citet{Mannucci_06} find that about 50 \% of SNe Ia explode soon after their stellar birth, and further works reveal that the DTD is proportional to $t^{-1}$ with its peak at around 0.1 Gyr \citep{Totani_08, Maoz_10}. Therefore, the phenomenon that numerous SNe Ia sequentially explode after the bursting explosions of SNe II would be realized if a star cluster was formed prior to the formation of NGC 1718.

Let's calculate the mass of star cluster as a source of numerous SN Ia ejecta, that may or may not finally become field stars after its disintegration. From the assumptions that 5\% of stars with the mass range of 3-8 \ms eventually become SNe Ia \citep{Tsujimoto_12} and the fraction of prompt SNe Ia is 50\% \citep{Mannucci_06}, the mass of $\sim 4\times10^4$\ms is deduced. This cluster yields more than 200 SN II explosions in total within a few $10^7$yrs prior to the commencement  of SN Ia explosions, which is likely to provide sufficient  energy to expel the ISM surrounding the star cluster. The resultant structure must be identical to a HI hole that can be ubiquitously seen in the present-day LMC disk \citep{Kim_99}. Then finally, a low-metallicity gas from the SMC rains down on this spot, which is followed by sequential SN Ia explosions. 
Accordingly, our proposed scenario is summarized as follows;

1.At the beginning, a star cluster with the mass of 

\hspace{0.3cm}$\sim 4\times10^4$\ms is formed.

2.Subsequently, a busting SNe II explosions expel  the

\hspace{0.3cm}surrounding ISM of this cluster, and make a HI hole.

3.Onto this HI hole, a gas disrupted from the SMC  

\hspace{0.3cm}with a metallicity of [Fe/H]$<$-1.3 accretes.

4.Sequential prompt SNe Ia start to explode and the 

\hspace{0.3cm}multiple ejecta of SNe Ia merge and mix with the  

\hspace{0.3cm}new ISM supplied by the accreting metal-poor gas.  

\hspace{0.3cm}Finally, NGC 1718 is formed from its mixed gas.

\noindent This is the  formation process of NGC 1718
 that took place about 2 Gyr ago in the LMC.

\section{SN Ia-like abundances of Fe-peak elements}

\begin{figure}[t]
\vspace{0.2cm}
\begin{center}
\includegraphics[width=7cm,clip=true]{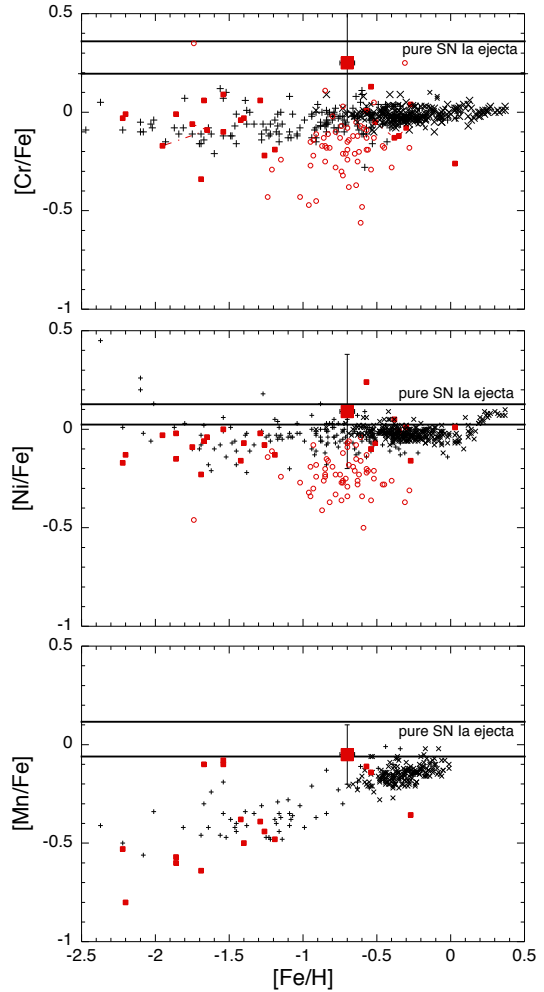}
\end{center}
\vspace{0.3cm}
\caption{The observed [Fe-peak elements/Fe] ratios of NGC 1718 \citep[red filled square:][]{Colucci_12} together with the stars in the LMC and the Galaxy, compared with the theoretical nucleosynthesis ratios in SNe Ia \citep[solid lines:][]{Iwamoto_99}. For the observed data of the LMC, the GCs and field stars are denoted by filled small square (Johnson et al.~2006; Mucciarelli et al.~2008, 2010, 2012; Colucci et al.~2012) and open circles \citep{Pompeia_08}, respectively. For the Galaxy data, disk stars and halo stars are denoted by crosses \citep{Reddy_03, Bensby_05} and pluses \citep{Gratton_03}, respectively. Two theoretical lines are the results from two different models for SNe Ia \citep[see][]{Iwamoto_99}.
}
\end{figure}

The unusual elemental feature of NGC 1718 is seen in not only [Mg/Fe] but also the ratios of Fe-peak elements (Cr, Mn, Ni) to Fe. Figure 3 shows the observed [Cr, Ni, Mn/Fe] ratios of NGC 1718, compared with the LMC and the Galaxy data. We see higher ratios of NGC 1718 than those of other field/cluster stars, in particular for [Cr/Fe]. The reason why these ratios are unusually higher and the deviation of [Cr/Fe] is most significant can be well understood if we compare the observed data with the theoretical nucleosynthesis result of SNe Ia. To consider an uncertainty in nucleosynthesis calculations, the predicted ratios by two models \citep[delayed-detonation models: WDD1, WDD2 in][]{Iwamoto_99} are attached to each panel of Figure 3. We see a good coincidence between the observed ratio for NGC 1718 and the predicted range given by two nucleosynthesis models. 
This is a compelling evidence supporting our scenario, because the  [Cr, Ni, Mn/Fe] ratios inside the ejecta 
that eventually gives birth to NGC 1718 should basically retain the ratios predicted by nucleosynthesis in SN Ia, with a only small difference by $\sim$0.02-0.1 dex from a pure ejecta case.

\section{AGB-like abundances of light-odd elements}

Light odd-elements, Na and Al,  are synthesized in asymptotic giant branch (AGB) stars, with a production peak at a $\sim$5 \ms AGB star \citep[e.g.,][]{Fenner_04, Karakas_07}. Since the lifetime of 5 \ms star is $\sim$0.1 Gyr, which is nearly equivalent to the major delay time for prompt SNe Ia, the ejecta of prompt SNe Ia might be  unavoidably  contaminated by the release of Na and Al from mass-losing AGB stars. Indeed, the observed Na and Al abundances of NGC 1718 are not low, i.e., [Na/Fe]=+0.05$\pm$0.22, [Al/Fe]=+0.15$\pm$0.30 \citep{Colucci_12}, though these elements are produced little in SNe Ia \citep{Iwamoto_99}. 

Here we try to make a quantitative estimate of the contribution from AGB stars. Roughly, $\sim$ 20 AGB ejecta per one SN Ia are expected from a simple yet reasonable assumption of SNe Ia/AGB $\sim$0.05 \%. The AGB yields are updated by several authors \citep[e.g.,][]{Fenner_04, Karakas_07, Izzard_07}, and these yields are recently examined to explain the origin of Na-O and Al-Mg anticorrelations observed in the Galactic GCs \citep{Marcolini_09, Sanchez_12}. If we adopt the mean Na yield averaged over 4-7 \ms AGB stars of 7.7$\times 10^{-4}$ \ms for $Z$=0.004 model by \citet{Karakas_07}, the ejecta of SNe Ia associated with 20 AGB ejecta results in [Na/Fe]=-0.03, which is in good agreement with the observed ratio. For the Al yield, if we adopt the high yield of 2$\times 10^{-3}$ \ms for instance, which is close to the value of 3.5$\times10^{-3}$ that nicely explains the Al-Mg anticorrelation \citep{Marcolini_09}, we deduce [Al/Fe] =+0.15  for the SN Ia+AGBs ejecta yielding [Mg/Fe]=-0.9.  

\section{Discussion and conclusions}

NGC 1718 is the only GC that has been observed so far to show a very low [Mg/Fe]. However, since a large amount of gas as much as  $\sim10^6 -10^8$\ms is predicted to be accreted onto the LMC from the SMC about 2 Gyr ago \citep{Bekki_07, Diaz_12},  there will be a potentiality of detection of  field stars whose chemical features are identical to NGC 1718 in the LMC disk. The place where these stars populate is likely to be off the central region like the location of NGC 1718, because both the observed distribution of giant HI holes and the predicted spots of accretion are outside the central part. Note that the observed data for the LMC field in Figure 1 is the inner disk sample \citep{Pompeia_08}. Therefore, we propose that future spectroscopic observations for the outer disk will detect the stars with unusually low [Mg/Fe] ratios. The level of low Mg/Fe for these individual stars must vary with various [Fe/H] owing to (i) the different number of SN Ia explosions in local regions  and 
(ii) the wide range of metallicity of the  accreted metal-poor gas. 
 
According to the prediction by \citet{Bekki_07}, the LMC has experienced another accretion event associated with a large amount of metal-poor gas up to $10^8$ \ms from the SMC about 0.2 Gyr ago. This prediction leads to the possible presence of young GCs and/or field stars exhibiting a low [Mg/Fe] ratio in the proposed scenario. One candidate at the moment is the young cluster NGC 1866 with [Fe/H]=-0.27 and its age of 0.1-0.5 Gyr, which exhibits [Mg/Fe]=-0.27$\pm$0.20 \citep{Colucci_12}. It should be, however, noted that other [$\alpha$/Fe] ratios in this cluster are larger than 0.

Except for NGC 1718, two stars are found in the Carina dSph to have very low [Mg/Fe] ratios less than -0.7  \citep{Lemasle_12}. In addition to them, \citet{Venn_12} find the star ({\it Car}-612) with [Mg/Fe]=-0.5$\pm$0.16 at [Fe/H]=-1.3 in the Carina, and discuss its origin as a pocket of SN Ia enriched gas \cite[see also][]{Marcolini_09, Sanchez_12}. Note that this star's [Mg/Fe] is measured to be -0.9 by \citet{Koch_08}. The presence of these stars may suggest that in dwarf galaxies, stars formed from the ejecta of SNe Ia can retain its relic thanks to the low-metallicity environment such as [Fe/H]$\sim$ -1.3, though iron-peak elements of {\it Car}-612 do not show SN Ia-like abundances as discussed in \S 4 such that  [Cr/Fe], [Mn/Fe], [Ni/Fe] are -0.20, -0.51, -0.46, respectively. Since the Galactic halo is considered to be  formed from the ancient destruction of dwarf galaxies, there may exist halo stars with an unusually low [Mg/Fe] originated from the SNe Ia ejecta in dwarf galaxies. The newly identified halo stars by \citet{Ivans_03}, i.e., BD +80$^\circ$245, G4-36, and CS 22966-043 with [Mg/Fe]=-0.22, -0.19, and -0.65, respectively, are promising candidates. 

Here we present a bigger picture for the GC formation; past gas accretion events triggered the formation of young and intermediate-age GCs in the LMC. This view could provide a clue to  the origin of the  age-gap problem, i.e., the observed fact that almost all GCs in the LMC are either very old (like the Galactic GCs) or younger than a few Gyr \citep[e.g.,][]{Costa_91}. Since most GCs are formed from the mixture of an enriched ISM and an accreted metal-poor gas, their chemical abundances are predicted to be similar to those in the ISM at their birth epoch. It is already claimed that the strong LMC-SMC interaction is responsible for the onset of GC formation in the LMC disk \citep{Bekki_04}. Its outcome, gas accretion, could also be a driver of the GC formation in the LMC.

In this {\it Letter}, we have shown that an unusual low [Mg/Fe] ratio recently found for one intermediate-age GC in the LMC enables us to assess its unique origin. Since its extremely low ratio ($\leq$-0.6) is outside any observed  Al-Mg anticorrelations ($>$-0.3) as well as by the prediction from nucleosynthesis calculations on any SNe II ($>$-0.2, though there is room for possibility of lowering this limit in some low-mass SN II models), a birth place of this GC is narrowed down to the ejecta of SNe Ia. However, in general, an already chemically enriched ISM, e.g., with [Fe/H]$\sim$-0.5, eventually changes the chemical abundances inside the ejecta from those determined by the chemical pollution by SN explosions to those similar to the ISM, because the total amount of metals ejected by SNe is much smaller than that contained in the ISM. Therefore, the formation of the GCs showing characteristics of the SN nucleosynthesis at a moderate metallicity is possible only under two physical conditions that  (i) there are some local regions devoid of HI gas in the  galaxy, and that  (ii) the galaxy experiences accretion of  very metal-poor gas. The LMC is one of galaxies where these two conditions can be satisfied. 

\acknowledgements
The authors wish to thank the referee Brad Gibson for his valuable comments that helped improve the paper. This research was supported by the Graduate University for Advanced Studies (SOKENDAI). TT is assisted in part by Grant-in-Aid for Scientific Research (21540246) of the Japanese Ministry of Education, Culture, Sports, Science and Technology.

\end{document}